\documentclass[prl,twocolumn,amsmath,amssymb,superscriptaddress]{revtex4-1}
\pdfoutput=1

\usepackage[pdftex]{graphicx}
\usepackage[pdftex,colorlinks=true,urlcolor=blue,linkcolor=black]{hyperref} 

\usepackage{graphicx}
\usepackage{dcolumn}
\usepackage{bm}
\usepackage{datetime}
\usepackage{color}
\usepackage{float}
\usepackage{enumitem}
\setitemize{leftmargin=*} 

\usepackage{ulem}				

\newcommand{\bp}{\textbf{p}}
\newcommand{\bk}{\textbf{k}}

\newcommand{\bx}{\textbf{x}}
\newcommand{\sks}{\sigma^k_s}
\newcommand{\skn}{\sigma^k_n}
\newcommand{\bnabla}{\vec{\nabla}}

\newcommand{\braket}[1]{\langle #1 \rangle}	

\begin{document}

\title{Direct imaging of the order parameter of an atomic superfluid using matterwave optics}

\author{Puneet A. Murthy}
\email{murthyp@phys.ethz.ch}
\affiliation{Physics Institute, Heidelberg University, Germany}
\affiliation{Institute for Quantum Electronics, ETH Z{\"u}rich, Switzerland}

\author{Selim Jochim}
\affiliation{Physics Institute, Heidelberg University, Germany}


\begin{abstract}
We propose a method to directly measure the complex phase distribution, superfluid density and velocity field in an ultracold atomic superfluid. The method consists of  mapping the momentum distribution of the gas to real space using matterwave focusing, and manipulating the amplitude and phase by means of tailor made optical potentials. This makes it possible to find analogues of well-known techniques in optical microscopy such as Zernike phase contrast imaging, dark field imaging and schlieren imaging. Applying these ideas directly at the level of the macroscopic wavefunction of the superfluid will allow visualization of interesting effects such as phase fluctuations and topological defects, and enable measurements of transport properties such as vorticity.
\end{abstract}

\pacs{Valid PACS appear here}
\maketitle

Superfluidity is an intriguing phenomenon where quantum mechanical effects emerge at macroscopic scales. With the advent of ultracold Bose and Fermi gases, an excellent experimental platform has emerged for studying superfluids in a variety of settings. The unprecedented level of tunability in these systems has allowed the measurement of physical observables that are difficult to access in condensed matter systems such as $^4$He. Interference and time-of-flight experiments in both Bose and Fermi systems have provided striking visualizations of long range coherence \cite{Bloch2000,Andrews1997}, topological defects such as vortices \cite{Zwierlein2005} and solitons \cite{Denschlag2000}, and correlation functions \cite{Hadzibabic2006,Murthy2015,Langen2013}. Currently, a range of fundamental questions are being pursued under the broad themes of lower dimensional systems - such as Berezinskii--Kosterlitz--Thouless transition in 2D; non-equilibrium phenomena such as the Kibble Zurek mechanism \cite{Navon2015,Aidelsburger2017}, superfluid turbulence \cite{Navon2016,Henn2009} and non-thermal fixed points \cite{Berges2008}. To address these questions, robust experimental tools are required that allow direct access to the key observables, and consequently simplify the interpretation of experimental data.
\begin{figure}[t!]	
\includegraphics[]{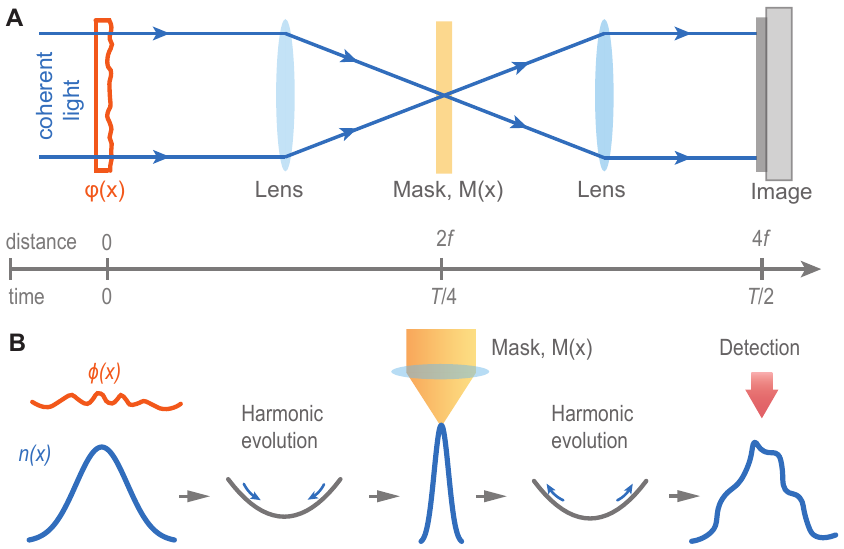}
	\caption{\textbf{Microscopy of the matterwave field.} \textbf{A} Illustration of the 4\textit{f} imaging setup in optics, which utilizes the Fourier transforming property of a lens. An object modifies an incident coherent light field, which propagates through a series of lenses. At the first Fourier plane at a distance $z=2f$, a mask function $M(\bx)$ is imposed on the field followed by propagation through a second lens. \textbf{B} An analogous method performed with an ultracold atomic superfluid. Using a matterwave lens that consists of an evolution in a harmonic trap for a quarter period $t=T/4$, the system is transformed to Fourier space where a Mask function is applied for a short duration by means of spatially engineered optical potentials. After evolution for another $T/4$, the spatial density distribution of the gas is imaged.} 
	\label{fig:method}
\end{figure}

Here, we propose a method to directly measure different components of the order parameter \cite{Leggett2001,Yang1962} of a trapped atomic superfluid, $\Psi(\bx,t) = \sqrt{n_s(\bx,t)}\exp(i\phi(\bx,t))$,  by imaging the density $n_s$, the local phase $\phi$ and the local gradients of the field. Our method is analogous to $4f$-imaging in optical microscopy \cite{Saleh1991} (see Fig.\ref{fig:method} A) from which several well known techniques such as dark-field, phase-contrast and schlieren imaging can be derived. While the optical versions of these techniques have been widely used in cold atom physics to measure density distributions, here we show that these ideas can be applied directly on the macroscopic wavefunction thus providing access to the interesting properties of superfluids.

The crux of the method lies in the Fourier transforming property of a harmonic potential, i.e. the evolution of a quantum state in a harmonic trap for a quarter of its period amounts to performing a Fourier transform ($\mathcal{F}$) of the initial wavefunction \cite{Murthy2014}. In cold atom experiments, the measurement of momentum distribution using harmonic potentials has been demonstrated for both 1D \cite{Jacqmin2012,Shvarchuck2002,VanAmerongen2008} and 2D \cite{Tung2010,Murthy2014} systems by means of external optical and magnetic potentials. To describe our method, we consider a gas of bosons trapped in an external potential and whose state is described by the field $\Psi(\bx,t=0)$. The gas is suddenly released to expand ballistically in a shallow harmonic potential for a quarter period $T/4 = \pi/2\omega$ (Fig.\ref{fig:method} B). At $t=T/4$, the different momentum components of the initial state separate spatially, and the field and density are given by
\begin{equation}\label{eq:T4}
\begin{split}
{{\Psi}}(\tilde{\bx}, T/4) & = \mathcal{F}[\Psi(\bx, 0)] \equiv {\Psi}({\bk}, 0), \\
{{n}}(\tilde{\bx}, T/4) & = \tilde{n}(\bk, 0),
\end{split}
\end{equation}
where $\hbar{\bp} = m\omega\tilde{\bx}$ is the mapping between initial momentum and position $\tilde{\bx}$ at $t = T/4$, and $\mathcal{F}$ denotes a 2D Fourier transformation. The main requirement for this mapping to work is that the interparticle collisions during the $T/4$ evolution are negligible. This is readily achieved in both 1D and 2D systems, since the interaction energy is released primarily along the tightly confining direction leading to a rapid quench of interactions \cite{Murthy2014}. 

At $t = T/4$, a spatial mask $M(\tilde{\bx}) = A_\textrm{m}(\tilde{\bx})e^{i\varphi_\textrm{m}(\tilde{\bx})}$, with amplitude $A_\textrm{m}$ and phase $\varphi_\textrm{m}$, is imprinted on the field. The field is subsequently inverse Fourier transformed by yet another $T/4$ evolution which returns it to its original configuration but with mirrored spatial coordinates. The modified field at $t=T/2$ is then
\begin{equation}\label{eq:T2}
\Psi(\bx,T/2) = \mathcal{F}^{-1}[{M}(\tilde{\bx}).{\Psi}(\tilde{\bx},T/4)].
\end{equation}

Experimentally, these masks can be created with tailor made optical patterns generated using spatial light modulators (Liquid crystal phase modulators (LCOS) or Digital Micromirror Devices (DMD)), which have been implemented in several cold atom experiments in recent years \cite{Gaunt2013,Labuhn2016,Barredo2017}. To apply binary amplitude masks, we propose using light pulses that are resonant to the atomic hyperfine transitions. Atoms at certain momenta can be selectively filtered out by engineering the spatial form of the resonant beam. On the other hand, to modify only the phase, we propose using optical potentials that are far detuned from the hyperfine transitions. Such far-detuned potentials have been previously employed to imprint vortices and solitons \cite{Denschlag2000}. For a given far-detuned spatial potential $V(\bx)$, the phase imprinted on the field is 
\begin{equation}\label{eq:imprint}
\varphi_\textrm{m}(\tilde{\bx}) = -\frac{i}{\hbar} V(\tilde{\bx})\delta t,
\end{equation}
where $\delta t$ is the duration of the pulse. The required phase shift can achieved at much shorter time scales than the focusing time $T/4$ by tuning the overall intensity of the optical beam. We now proceed to separately outline the schemes required to measure the different components of the superfluid order parameter.\\

\begin{figure}[t!]	
	\includegraphics[]{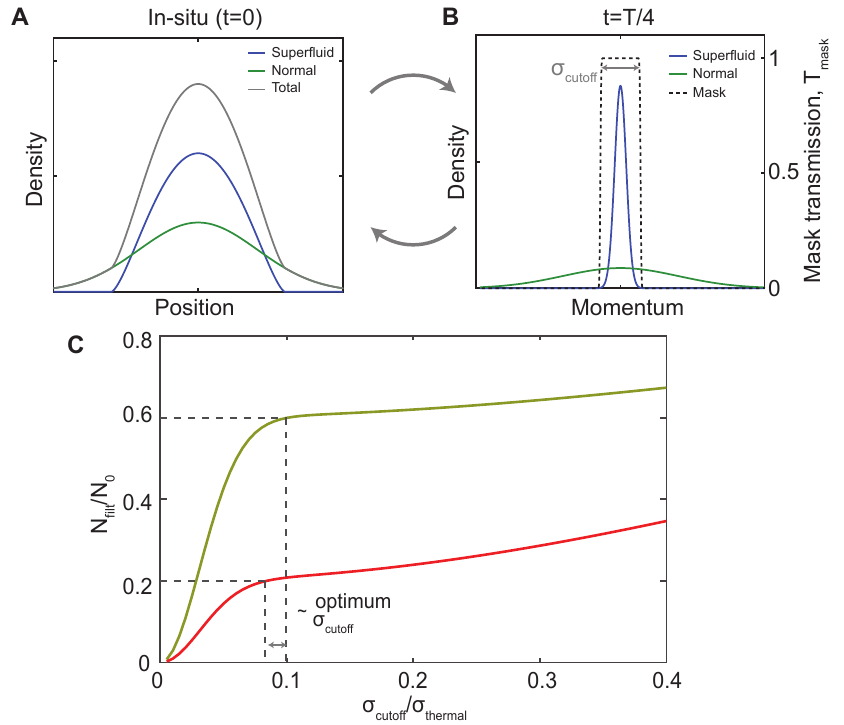}
	\caption{\textbf{Estimating the superfluid density.} \textbf{A.} In-situ density distribution of a harmonically trapped atomic superfluid which shows coexisting superfluid (blue) and normal (green) components. \textbf{B.} In momentum space ($t=T/4$), the superfluid occupies the low-$k$ modes while the normal gas has a broad distribution. Here, an absorptive mask (dashed line) with a cutoff scale $\sigma_\textrm{cutoff}$ is applied using a resonant laser beam, which removes the normal fraction. At $t=T/2$, the remaining superfluid density is measured. \textbf{C} The total fraction of unfiltered atoms $N_\textrm{filt}/N_0$ plotted as a function of $\sigma_\textrm{cutoff}$ scaled by the Boltzmann width $\sigma^k_{\textrm{n}}$ for two values of superfluid fraction $\xi=0.6$ (yellow) and $0.2$ (red). The curves show a clear shoulder as they cross the respective values of $\xi$ (dashed lines), which can be used to estimate the optimum  $\sigma_\textrm{cutoff}$.}  
	\label{fig:density}
\end{figure} 

\noindent
\textbf{Superfluid density.} We start by discussing a scheme to estimate the superfluid density. According to the two-fluid model, a system at a finite temperature below $T_c$ consists of a superfluid component $n_s = |\Psi|^2$ and an incoherent normal component $n_\textrm{n}$ \cite{Leggett2001}. In space, the two components coexist, as shown in Fig.\ref{fig:density} A, and the in-situ density is
\begin{equation}\label{n0}
n(\bx,t=0) = n_\textrm{s}(\bx,0) + n_\textrm{n}(\bx,0).
\end{equation}
Therefore extracting the superfluid density from the total in-situ profile is not a simple task. However, in momentum space the two components have vastly different distributions (see Fig.\ref{fig:density} B) owing to the large difference in coherence lengths between them. In general, the superfluid component occupies the low-lying momenta with width $\sks \sim 1/R_\mathrm{TF}$, where $R_\mathrm{TF}$ is the Thomas--Fermi radius. The thermal component on the other hand follows a broad Boltzmann distribution with width $\skn \sim 1/\lambda_T$ determined by the the thermal de-Broglie wavelength $\lambda_T$. In typical experiments, at low enough temperatures, the Thomas--Fermi radius is of the order of $100\,\mu$m and the thermal de-Broglie wavelength is $\sim 1\,\mu$m. Hence, the ratio of the two $k-$space widths can be of the order of $\skn/\sks = R_\mathrm{TF}/\lambda_\textrm{T} \sim 10-100$. Such a large separation of scales between the two components makes it possible to apply filtering operations on only one of them. To measure the superfluid density, we use an amplitude mask at $t=T/4$, represented by dashed lines in Fig.\ref{fig:density} B, that \textit{transmits} only the low$-k$ modes while discarding the thermal fraction. As the system is brought back to position space after another $T/4$ evolution, the measured density mainly consists of the initial superfluid density. 
\begin{equation}\label{eq:nT2}
n(\bx,T/2) \approx n_\textrm{SF}(-\bx,0).
\end{equation}

Experimentally, a low-pass filter is implemented using a spatially engineered resonant laser beam that is turned on for a short duration at $T/4$. Although quite simple in principle, the method's efficacy depends to some extent on the choice of the momentum-space cutoff $\sigma_\textrm{cutoff}$ of the filter. At a qualitative level, the measurement will generally be more accurate at lower temperatures where the ratio of the two widths is larger. At a given temperature, however, there exists an optimum cutoff scale which results in the smallest error in estimating the superfluid density. For a strongly bimodal momentum distribution, the optimum cutoff can be found by analyzing the total particle number as a function of $\sigma_\textrm{cutoff}$. To illustrate this, we consider the case where both components follow gaussian distributions with a width factor $\skn/\sks = 20$. It is quite straightforward to then show that the particle number after filtering $N_\textrm{filt}$ has the form,
\begin{equation}\label{eq:Nfilt}
\frac{N_\textrm{filt}}{N_0} = \alpha. \mathrm{erf}^2\left(\frac{\sigma_\textrm{cutoff}}{\sigma^k_\textrm{SF}}\right) + (1 - \alpha).\mathrm{erf}^2\left(\frac{\sigma_\textrm{cutoff}}{\sigma^k_\textrm{T}}\right),
\end{equation}
where $N_0$ is the total unfiltered particle number, $\alpha$ is the superfluid fraction and $\mathrm{erf}$ denotes the Gauss error function. In Fig. \ref{fig:density} C, we show the variation of the particle number after filtering as a function of $\sigma_\textrm{cutoff}$ for two different values of superfluid fraction: $\alpha = 0.6$, $0.2$. We note that both curves exhibit a clear shoulder feature where the filtered atomic fraction approaches the superfluid fraction and therefore the shoulder position can be used as a reliable proxy to determine the optimum $\sigma_\textrm{cutoff}$. We note that a similar shoulder should exist for any  typical distribution that has a large enough separation of scales.\\

\noindent
\textbf{Local Phase.} We now describe a scheme to measure the complex phase $\phi(\bx)$ of the superfluid. The phase is fundamental to the theory of superfluidity, as it describes the hallmark property that is long-range spatial coherence as well as the elementary excitations of the system such as phase fluctuations and topological defects (vortices and solitons) \cite{Bloch2008}. Absolute phase information is typically lost in experiments as only densities can be measured. Phase differences, however, can be converted to density variations using interferometric methods provided a suitable phase reference is available, either using a separately prepared coherent sample \cite{Polkovnikov2006,Niu2006,Hadzibabic2006} or by beam splitter operations \cite{Berrada2016}. 

Here we propose that, in analogy to Zernike phase contrast method \cite{Zernike1942}, the \textit{Bose-condensed} $k=0$ component of the system itself can be used as a phase reference, with respect to which the phase fluctuations are defined.
\begin{figure}[t!]	
	\includegraphics[]{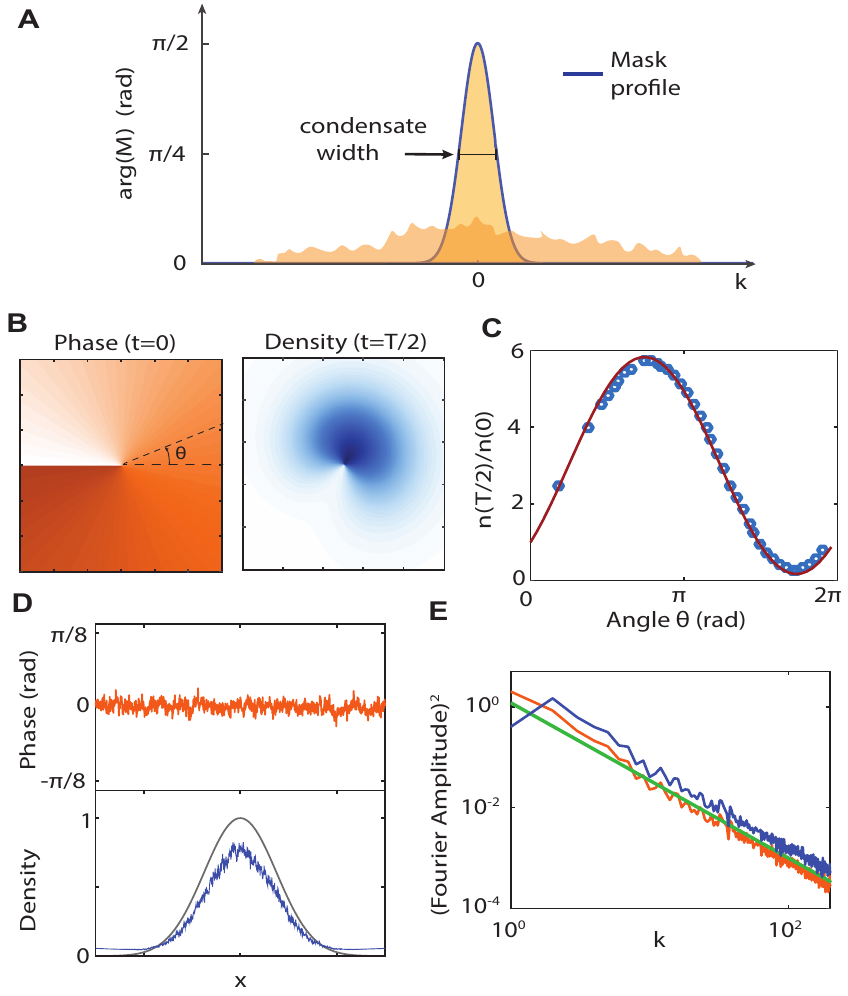}
	\caption{\textbf{Imaging the complex phase.} \textbf{A} In analogy to Zernike phase contrast microscopy, the mask function selectively imprints a $\pi/2$ phase shift on the $k=0$ condensate peak (yellow). At $t=T/2$, the density of the cloud then reflects the initial spatial phase $\phi(\bx)$. \textbf{B} The phase of a topological vortex (left panel) becomes visible as a winding of the density (lower panel) at $T/2$. \textbf{C} The scaled density $n(x, T/2)/n(x,0)$ at a fixed radius as a function of $\theta$ shows a sinusoidal dependence. \textbf{D} Random phase fluctuations with algebraically decaying Fourier spectrum $f(k) \sim k^{-1.5}$ (upper panel), which become measurable as density fluctuations (lower pane, blue line). \textbf{E}. The Fourier spectrum of the initial phase fluctuations (orange), the final density fluctuations at $T/2$ (blue) and the reference curve $f \sim k^{-1.5}$, show good agreement.} 
	\label{fig:phase}
\end{figure}
The complex field of the superfluid can be described as $\Psi(\bx) = \sqrt{n_0}e^{i(\phi_0 + \phi(\bx))} = \Psi_0(\bx)e^{i\phi(\bx)}$, where $\Psi_0$ is the wave-function of the condensate with constant phase $\phi_0$. The absolute value of the $\phi_0$ is inconsequential for physical properties of the system. The constant and fluctuating components of the field is written as \cite{Ketterle1999}, 
\begin{equation}\label{eq:phase1}
\Psi = \Psi_0 e^{i\phi(\bx)} = \Psi_0 + \delta\Psi. 
\end{equation}
To image the complex phase $\phi(\bx)$, we use a mask function that imprints a $\pi/2$ phase shift selectively on the Bose-condensed \textit{DC} component while leaving the rest of the system unaltered, as shown in Fig.\ref{fig:phase}A. Experimentally, this is achieved by a tightly focused off-resonant laser beam positioned at $k=0$, and having the same width ($\sks \sim 1/R_\mathrm{TF}$) as the $k$-space width of the condensate. The field at $t=T/2$ is the given by
\begin{equation}\label{eq:phase2}
\begin{split}
\Psi(\bx, T/2) & = \Psi_0 e^{i\pi/2} + \delta\Psi = \Psi_0[e^{i\pi/2} + e^{i\phi(\bx)} - 1].
\end{split}
\end{equation}

The corresponding density is $n(\bx,T/2) = \Psi^\dagger\Psi = n_0[3 - 2\sqrt{2}\cos(\phi(\bx) - \pi/4)]$, which shows a sinusoidal dependence on the spatial phase. For small phase shifts, we find the approximate density
\begin{equation}\label{eq:phase3}
n(\bx,T/2) \approx n_0[1 + 2\phi(\bx)]
\end{equation}
is directly proportional to the phase. 

We illustrate this scheme using two examples: a topological vortex and random phase fluctuations with a power law decaying spectrum. The examples are physically relevant for studies on 2D superfluidity described by Berezinskii--Kosterlitz--Thouless (BKT) theory. The computation consists of Fourier transforming the initial field, multiplying a factor $e^{i\varphi_\textrm{mask}}$ and Fourier transforming back. In Fig.\ref{fig:phase} B and C, we show the case of a topological vortex $\phi(\bx) = m\theta$ (left panel in Fig.\ref{fig:phase}B), where $m = \pm 1$ is the winding number. We see that, due to the $\pi/2$ phase shift on the $k=0$ component, the winding of the initial \textit{phase} at $t=0$ is converted a winding of the \textit{density} at $t=T/2$. In Fig.\ref{fig:phase} C, we show the scaled density $n(\bx,T/2)/n(\bx,0)$ as a function of the azimuthal angle $\theta$ for a fixed radius, which shows the expected sinusoidal dependence on the phase.

In Fig.\ref{fig:phase} D and E, we show the example of phase fluctuations in the superfluid. The initial phase chosen here (Fig.\ref{fig:phase} D, upper panel), contains random fluctuations with a Fourier spectrum that decays algebraically according to $f(k) \sim k^{-2 + \eta}$, with $\eta = 0.5$. This corresponds to the spatial first order correlation function having the form $g_1(r) = \braket{e^{i(\phi(r) - \phi(0))}} \sim r^{-\eta}$. At $t=T/2$, the random fluctuations in the initial phase are converted to density fluctuations (Fig.\ref{fig:phase} D lower panel), which are simple to measure. In Fig.\ref{fig:phase} E, we show the Fourier spectrum of the scaled density at $T/2$ (blue), the initial phase (orange) and a reference curve $f(k) \sim k^{-1.5}$ (green), which agree with each other. Since the spectrum directly relates to the phase coherence function, this implies that density-density correlations at $T/2$ directly relates to $g_1(r)$ at $t=0$. Power law behavior of phase coherence is found in several phenomena, notably near critical transitions and non-equilibrium systems. Measuring the exponents is therefore relevant for current studies both in lower dimensions \cite{Boettcher2016} and non-equilibrium systems \cite{Navon2016}.\\ 

\begin{figure}[t!]	
	\includegraphics[]{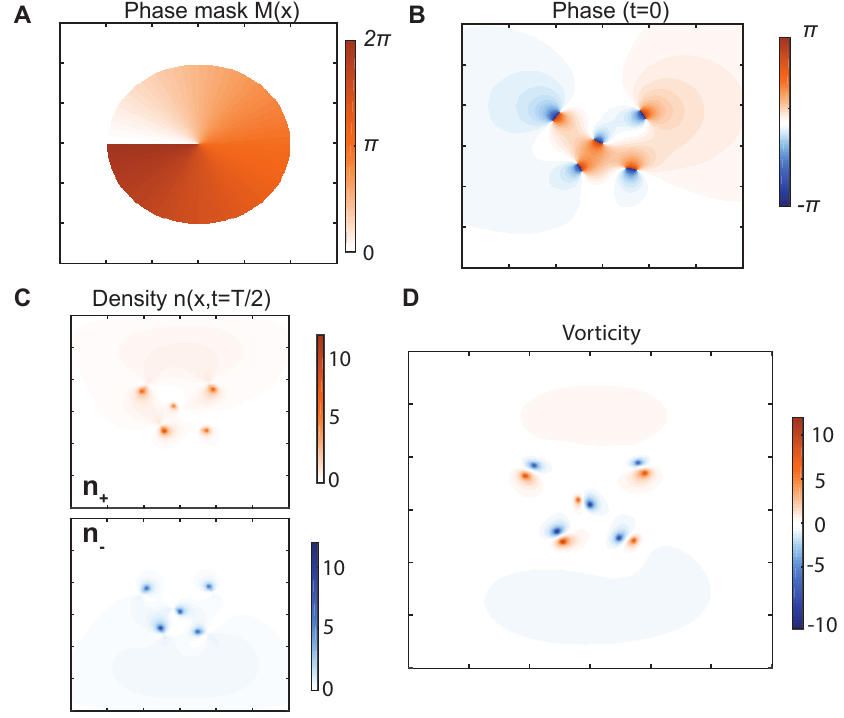}
	\caption{\textbf{Vorticity.} \textbf{A} A spiral phase mask $\arg(M) = \pm\theta$ is applied at $t=T/4$, which leads to two density distributions $n^+$ and $n^-$ at $t=T/2$ accordingly. \textbf{B} Example of a distribution of topological vortex--antivortex pairs. \textbf{C} We show $n_+$ (upper panel) and $n_-$ (lower panel) corresponding to $m=+1$ and $m=-1$. In each of the two images, only the vortices with the circulation parallel to the mask become visible at $t=T/2$. \textbf{D} The vorticity of the superfluid can be extracted directly from the difference between the two densities $n_+ - n_-$, normalized by the reference superfluid density.} 
	\label{fig:spiral}
\end{figure}

\noindent
\textbf{Local velocity and vorticity.} Transport properties of superfluids are essentially encoded in the derivatives of the phase. The two relevant quantities in this regard are the superfluid velocity $\mathbf{v_s}(\bx) = (\hbar/2m)\bnabla \phi(\bx)$, and the vorticity $\mathbf{\omega} = (\hbar/m)\bnabla \times {\bnabla} \phi(\bx)$. The latter characterizes rotationality in superfluids and is non-zero only in the presence of a phase singularity such as a vortex or flow around an obstacle. Both $v_s$ and $\vec{\omega}$ play an important role in the study of non-equilibrium phase transitions (eg. Kibble-Zurek mechanism) and superfluid hydrodynamics (eg. 2D turbulence).

To measure $v_s$ and $\vec{\omega}$, the distribution of phase gradients should be directly converted to a density distribution. This can be achieved by using a spiral mask $M(\tilde{\bx}) = e^{\pm i\theta}$ (see Fig.\ref{fig:spiral} A), which has the same form as the phase of a vortex except here it is applied in $k-$space. The main function of this mask is to imprint a $\pi$ phase difference between any two diametrically opposite points. 

In optics, such spiral phase plates have been used to enhance edge features of objects and characterize the orbital angular momentum of optical fields \cite{Lowenthal1967,Fuerhapter2005a,Juchtmans2016,Juchtmans2016a}. An important aspect here is that the direction of circulation of the spiral mask impacts the final field at $T/2$, as we will show below. To analyze the effect of the spiral mask, it is simpler to work with polar coordinates $(x,y) \rightarrow (r,\theta)$. The field at $t=T/2$ is given by
\begin{equation}\label{eq:spiral1}
\Psi(\textbf{r},T/2)  = \mathcal{F}^{-1}[e^{\pm i\theta_k}.\mathcal{F}[\Psi(\textbf{r})]] = \Psi(\textbf{r}) * \mathcal{F}(e^{\pm i\theta_k})(\textbf{r}),
\end{equation}
\noindent   
where $*$ is the convolution between functions. The Fourier transform of the phase mask is a radially symmetric function $ \mathcal{F}(e^{\pm i\theta})(\textbf{r}) = 2\pi i e^{\pm i\theta} \int_{0}^{k_\textrm{max}}dk k J_1(kr)$, where $k_\textrm{max}$ is a cutoff momentum upto which the mask is active. Following the analysis of \cite{Juchtmans2016}, the final field has the form 
\begin{equation}\label{eq:spiral2}
\Psi_{\pm}(\textbf{r},T/2) = C\left[\partial_r \Psi(\textbf{r},0) \pm \frac{i}{r} \partial_\theta \Psi^*(\textbf{r},0)  \right],
\end{equation} 
where $C = \pi e^{i(\theta + \pi)}\int dr' r'^2 \int_{0}^{k_\textrm{max}}{dk k J_1(kr)}$ is a normalization factor that depends on the cutoff momentum. It is evident that  depending on whether the mask circulation is clockwise or anti-clockwise, we get different expressions for the field. For each of the two cases ($\Psi_{+}$, $\Psi_{-}$), we can extract the corresponding density ($n_+$, $n_-$). The velocity and vorticity can be obtained from the sum and difference of the densities of the two components. From the sum of the components, 
\begin{equation}\label{eq:spiral3}
n_+ + n_-  = 2C^2[\partial_r \Psi \partial_r \Psi^* + \frac{1}{r^2}\partial_\theta \Psi \partial_\theta \Psi^*] = 2C^2 |\bnabla_r \Psi|^2, 
\end{equation}
\noindent
we obtain the absolute gradient of the field along the radial direction. For the case of irrotational flow and slowly varying density, this can be further reduced to obtain $n_+\,+\,n_- = 2C^2n_s(\bx,0)|\bnabla\phi(\bx)|^2$. The radial component superfluid velocity has the form,
\begin{equation}
|\mathbf{v}_s|_r \simeq \frac{\hbar}{4mC} {\left(\frac{n_+ + n_-}{n_s}\right)}^{1/2}. 
\end{equation}

On the other hand, the difference between the two components is sensitive to the angular variation of the phase gradient,
\begin{equation}\label{eq:spiral4}
\begin{split}
n_+ - n_- & = \frac{2 i C^2}{r} [\partial_r \Psi \partial_\theta\Psi^* - \partial_\theta\Psi \partial_r \Psi^*], \\
		& = {2C^2} (\bnabla \times \mathbf{J}),
\end{split}
\end{equation}
where $ \mathbf{J} = \frac{\hbar}{2m}(\Psi^* \bnabla \Psi - \Psi \bnabla \Psi^*) \simeq \frac{\hbar}{2m}n_s\mathbf{v_s} $ is the superfluid current. The normalization factor $C$ tends to unity if the size of the spiral phase plate in momentum space is larger than typical momentum scales of the system. Then, the vorticity of the superfluid can be obtain according to 
\begin{equation} \label{eq:vorticity}
\vec{\omega} = \frac{\hbar}{m}\bnabla \times \mathbf{v_s} \simeq \frac{n_+ - n_-}{n_s}.
\end{equation} 

In Fig.\ref{fig:spiral} B-D, we exemplify this scheme using a phase distribution consisting of vortex-antivortex pairs (upper panel, B). These dipoles are formed with two vortices of opposite circulation with spatial separation. They play a prominent role in the phenomenology of 2D superfluids as described by BKT theory. We perform the computation of the field and density at $t=T/2$ for both clockwise and anti-clockwise spiral masks. The corresponding densities $n_+$ and $n_-$ thus obtained are shown in \ref{fig:spiral} C. An interesting feature of this scheme is that at $t=T/2$, the density accumulates at the vortex core locations where the field gradient is the largest. Furthermore, we see that in each of these images, only the vortices with circulation parallel to the spiral mask become visible. The difference between the images gives us the local vorticity which shows sharp positive and negative peaks corresponding to vortices of opposite circulation (Fig.\ref{fig:spiral} D). 

Indeed, the spiral phase mask function is the two-dimensional extensition of the 1D Sign function,
\begin{equation}
S(k_x) = 
\begin{cases}\label{eq:sgn}
-i = e^{-i\frac{\pi}{2}} &\quad k_x <0 \\
i = e^{+i\frac{\pi}{2}} &\quad k_x > 0, 
\end{cases} 
\end{equation} 
which shifts the phase of the function by $\pm \pi/2$ for positive and negative momenta. Therefore the 1D version of this mask can also be used to study phase gradients and dislocations along one chosen direction, which may be particularly useful to study soliton physics in cold gases. Interestingly, the application of such a mask in $k-$space can also be represented in terms of Hilbert transformations which are prominently used in signal processing. The Hilbert transform of a function $f(x)$ is it's convolution with $\frac{1}{x}$, i.e $\mathcal{H}(f(x)) = -\frac{1}{\pi}f(x) * \frac{1}{x}$. Hilbert transforms are typically used to obtain the analytic representation of an arbitrary waveform. In the current scenario, the wavefunction of the superfluid at T/2, after passing through the Heaviside phase mask (Eq.\,\ref{eq:sgn}), can be written in terms of the the initial spatial wavefunction according to,
\begin{equation}
\Psi(-\textbf{r},T/2) =\mathcal{H}[\Psi(\textbf{r},0)] = \mathcal{F}^{-1}[-iS(k_x) . \mathcal{F}[\Psi(\textbf{r},0)]]. \nonumber
\end{equation}

To realize this experimentally, one of the key requirements is the simultaneous measurement of both $n_+$ and $n_-$ in each realization. For this, we propose the following scheme that exploits the availability of hyperfine states in atomic systems. Close to $t=T/4$, roughly half of the atoms are transferred to a higher lying hyperfine state using a microwave/radio-frequency $\pi$-pulse. The light for the phase mask is detuned (between the two transitions) such that atoms of one spin component are red-detuned while those of the other component are blue-detuned with respect to the laser. For the spiral mask, this translates to the two spin components experiencing opposite circulations. Subsequently at $T/2$, the density of each component is measured separately in each shot. 

We now turn to a concrete experimental example where these ideas could be realized. We consider a 2D gas of approximately $5\times 10^4$ $^{87}$Rb atoms in the ground state 5$S_{1/2}$, trapped in a highly anisotropic harmonic potential,  cooled to low temperatures $T \sim 100\,$nK and having a radial Thomas--Fermi radius of $R_\textrm{TF} = 50\,\mu$m.  The harmonic focusing potential is assumed to have a frequency $\omega = 2\pi \times 5\,$Hz and a focusing time $T/4 = 50\,$ms. There are three main requirements for the methods proposed above. (i) Sufficient spatial resolution at $T/4$, such that different $k$-components can be addressed using the optical masks. For the assumed frequency ($\omega = 2\pi \times 5\,$Hz), we find the $k$-space width of the Bose-condensed component to be $\sks = 2\pi \hbar /m\omega R_\textrm{TF} \approx 3\,\mu$m, which is resolvable with an objective ($\textrm{NA} \sim 0.2$). (ii) The phase imprinting operations should be performed at time scales much shorter than the atomic motion. For a laser beam with wavelength $\lambda = 1064\,$nm, we estimate that a beam power of $I=10\,$mW focused down to a spot of diameter $3\,\mu$m and duration $20\,\mu$s is sufficient to imprint a $\pi/2$ phase shift, which is experimentally feasible. (iii) For the hyperfine state transfer scheme suggested above, we use the manifolds $F=1$ and $F=2$, which are energetically separated by 6.8 GHz. A microwave $\pi$-pulse can be used to transfer approximately half of the atoms to the higher state in a short enough amount of time.

\noindent
\textbf{Magnification.} So far, we have considered the situation where the two matterwave lenses in the $4f$ setup are identical, i.e. the same trap frequencies for the two halves of the evolution ($0 < t < T/4$ and $T/4 < t < T/2$). In this scenario, the spatial scale at $t=T/2$ remains the same as the initial system, with only the coordinates being mirrored. However, depending on experimental requirements, it may be desirable to \textit{magnify} features in the system. For instance, the healing length of vortices might be smaller than the optical resolution, making it difficult to image vortices using the methods presented above. This can be solved by magnifying the matterwave field, which is simply achieved by having two different harmonic trap frequencies for the two halves of the evolution. The effective magnification is then given by $M = \omega_2/\omega_1$, where $\omega_1$ and $\omega_2$ are the trap frequencies for $0<t<T/4$ and $0<t<T/2$. This is analogous to using lenses with different focal lengths in an optical $4f$ setup. 

To summarize, we proposed a set of methods that enable direct imaging of the order parameter of an ultracold atomic superfluid. The individual tools suggested in this work rely on already established techniques, such as matterwave focusing, phase imprinting, spatially engineered potentials and hyperfine state transfer. The ideas presented here can be straightforwardly applied to superfluids in a variety of trap geometries such as optical lattices, box traps and annular traps. In the future, it will be exciting to build upon these ideas and develop tools that can be used to measure other fundamental properties of quantum systems related to entanglement and many-particle correlations \cite{Bergschneider2019}, where the mask acts on the wavefunction at the single particle level.

\noindent
\textbf{Acknowledgements.} The authors gratefully acknowledge Thomas Lompe, Igor Boettcher, Jean Dalibard, Philipp Preiss, Helmut Stroebel, Andrea Bergschneider and Thibault Chervy for insightful discussions and for providing comments on the manuscript.


%

\end{document}